\documentclass[epj]{webofc}
\usepackage[varg]{txfonts}   
\usepackage[mathletters]{ucs}

\woctitle{MENU 2013}
\begin{document}
\title{Low-energy constraints to $\alpha_s$}
%
%

\author{Xavier Garcia i Tormo\inst{1}\fnsep\thanks{\email{garcia@itp.unibe.ch}}
}

\institute{Albert Einstein Center for Fundamental Physics, Institut f\"ur Theoretische Physik, Universit\"at Bern,
  Sidlerstrasse 5, CH-3012 Bern, Switzerland }

\abstract{We briefly review some of the lower-energy constraints to the perturbative behaviour of the strong coupling $α_s$, with some emphasis on the determination coming from the energy between two static sources calculated on the lattice. }
\maketitle

The strong coupling, $α_s$, is the only free parameter of Quantum ChromoDynamics (QCD) in the massless quark limit. Its precise determination is of paramount importance for the study of processes that involve the strong interactions. Asymptotic freedom tells us that the coupling is perturbative at large energy. Its running with the energy scale is predicted by QCD, and encoded in the famous β function
\begin{equation}
μ\frac{dα_s(μ^2)}{dμ}=α_s(μ^2)β(α_s).
\end{equation}
$α_s(μ^2)$ is not continuous when crossing quark thresholds. Nowadays, the β function and the matching at quark thresholds are known to four loop order \cite{vanRitbergen:1997va,Czakon:2004bu,Schroder:2005hy,Chetyrkin:2005ia}. In the following we briefly overview some determinations of $α_s$ that are performed at perturbative but relatively low-energy scales. An advantage of a low-energy determination of $α_s$ is that, when its value is evolved to a higher-energy scale, like the $Z$-boson mass $M_Z$, the uncertainty shrinks, due to the logarithmic running. The downside of those determinations is that, since the value of $α_s$ is larger at lower scales, perturbative corrections are, in turn, larger and unknown higher-order terms could be important, and also that one needs to estimate or control non-perturbative effects more carefully. In any case, one wants to have $α_s$ determinations in the whole range of energies where perturbative QCD is valid, to obtain in this way a quantitative experimental test of asymptotic freedom.

\begin{figure}
\centering
\includegraphics[width=10cm,clip]{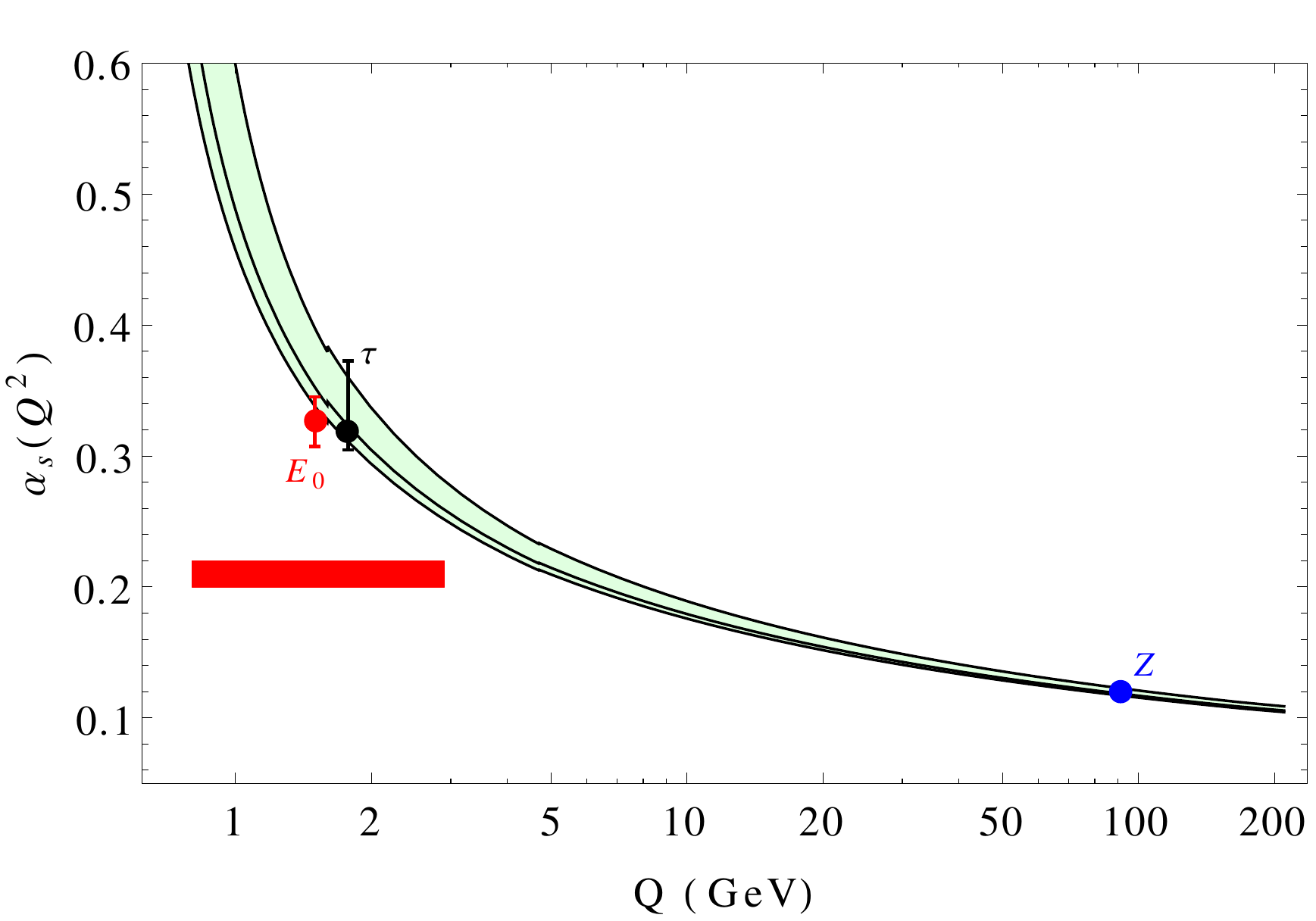}
\caption{$α_s(Q^2)$ as a function of the energy scale $Q$. The 4-loop running~\cite{Chetyrkin:2000yt} of the value obtained from data at the $Z$ pole~\cite{Beringer:1900zz} -blue point- is shown as the green band. For the $α_s$ value from τ decays -black point- we took the range spanned by several recent determinations \cite{Pich:2013sqa,Boito:2012cr,Abbas:2012fi}. The red band below the $α_s$ value from the static energy $E_0$ -red point- reflects the energy range that was used in this extraction~\cite{Bazavov:2012ka}.}
\label{fig:ZtauE0as}       
\end{figure}

A good and relatively clean way to obtain $α_s$ is to use processes that are inclusive hadronically, like the hadronic $Z$ decay rate, $R_Z:=Γ(Z\to\textrm{hadrons})/Γ(Z\to e^+e^-)$. The inclusive hadronic decay of the τ lepton allows us to determine $α_s$ at low energy~\cite{Braaten:1991qm}, i.e. at the scale of the τ mass $m_τ=1.78$~GeV. This observable has been extensively exploited in the literature during the years. Some of the main complications are related to how one organizes the perturbative expansion, i.e. how to treat higher-order corrections. Several treatments are present in the literature, and it is a long-standing discussion which method should give more accurate results. An adequate assessment of non-perturbative effects is also important, a recent analysis is given in Ref.~\cite{Boito:2012cr}. One can therefore discuss which should be the exact size of the error assigned to this result, but the fact remains that a comparison of this $α_s$ determination with the one coming from $Z$-pole data fits provides a striking confirmation of the predicted QCD running. This is illustrated in Fig.~\ref{fig:ZtauE0as}.

One can also obtain a determination of $α_s$ by comparing lattice data for the energy between two static sources in QCD, $E_0(r)$, and the corresponding perturbative expressions~\cite{Michael:1992nj,Booth:1992bm}. In this case, one can take advantage of recent progress in both the perturbative evaluation and the lattice computation of the static energy. Perturbatively the static energy is nowadays known at three loop, i.e. $\mathcal{O}(α_s^4)$, accuracy~\cite{Smirnov:2008pn,Anzai:2009tm,Smirnov:2009fh}, including also resummation of logarithmically enhanced terms at $\mathcal{O}(α_s^{4+n}\ln^nα_s)$ $(n\ge0)$~\cite{Brambilla:2006wp,Brambilla:2009bi}; a summary of all the currently known perturbative results can be found, for instance, in Ref.~\cite{Tormo:2013tha}. On the lattice side, the static energy with three light-quark flavors was recently computed in Ref.~\cite{Bazavov:2011nk}. A comparison of the two, and the corresponding extraction of $α_s$, was presented in Ref.~\cite{Bazavov:2012ka}; and corresponds to the red point in Fig.~\ref{fig:ZtauE0as}. One complication in this case is to know whether or not the current lattice data has really reached the purely perturbative regime, with precision enough to perform the extraction. It is difficult to undoubtedly state this point. In that sense, Ref.~\cite{Bazavov:2012ka} follows the idea that the agreement with lattice should improve when the perturbative order of the calculation is increased. This is found to happen, and the resulting perturbative curves can describe the lattice data quite well. In addition the result for $α_s$ is not very sensitive to the exact distance range that one uses in the analyses. These facts can be taken as an indication that one is indeed in the perturbative region. Further studies to verify this point are certainly warranted, though; we mention that Ref.~\cite{Leder:2011pz} concludes, in the two light-quark flavor case, that finer lattice spacings are needed for the extraction. An update of the analysis in Ref.~\cite{Bazavov:2012ka}, including lattice data with finer lattice spacings, and extensively addressing these questions is ongoing.

There are many other lattice determinations of $α_s$, which use different quantities and energy ranges. A relatively recent review of lattice results for $α_s$ is given in Ref.~\cite{McNeile:2013rga}. 

Another good way to extract $α_s$ at relatively low energies is to use ratios of quarkonium, $H$, decay widths. The complication here is that one needs to take into account the bound-state dynamics. The effective theory framework of Non-Relativistic QCD~\cite{Bodwin:1994jh} allows one to tackle with the problem. The best quantity for the $α_s$ extraction turns out to be the ratio $R_γ:=Γ(H\toγ+\textrm{hadrons})/Γ(H\to\textrm{hadrons})$, in the sense that it is the observable which is less sensitive to color-octet configurations and relativistic effects. An extraction for the bottomonium system, i.e. at the scale $M_{\Upsilon}=9.46$~GeV, was given in Ref.~\cite{Brambilla:2007cz}. A similar analysis for the charmonium system, i.e. at the scale $M_{J/\psi}=3.1$~GeV, is hindered by the fact that relativistic and octet corrections are more severe in this case.

There are several other good ways to determine $α_s$ at different energies, including parton distribution fits to deep inelastic scattering and hadron collider data, event shapes and jet rates in leptonic collisions, etcetera. Most of those results are collected and summarized in the Review of Particle Physics by the Particle Data Group (PDG)~\cite{Beringer:1900zz}, and in several other recent reviews on $α_s$ determinations, like Refs.~\cite{Pich:2013sqa,Altarelli:2013bpa}, the contents of which were helpful in preparing the present manuscript.

We finish by recalling that most, although not all, quantities entering the current (PDG) world average for $α_s$ are dominated by systematic errors of theoretical origin. These are many times difficult to precisely assess. In that sense, an increasing corroboration of the value of $α_s$, by extracting it from different independent quantities, and at different energy ranges, is both welcome and necessary.

\section*{Acknowledgments}
This work is supported by the Swiss National Science Foundation (SNF)
under the Sinergia grant number
CRSII2\underline{ }141847\underline{ }1.

\end{document}